\documentclass{article}
\usepackage{graphicx}
\usepackage{booktabs}
\usepackage{amsmath}
\usepackage{amsfonts}
\usepackage{amssymb}
\usepackage{url}
\usepackage{threeparttable}
\usepackage[margin=1in]{geometry}
\usepackage[table,xcdraw]{xcolor}

\title{Evaluating Speech-to-Text \(\times\) LLM \(\times\) Text-to-Speech Combinations for AI Interview Systems}
\author{
  Rumi Allbert \\ micro1 \\ \texttt{rumi@micro1.ai} \and
  Nima Yazdani \\ University of Southern California, micro1 \\ \texttt{nima@micro1.ai}
  \and 
  Ali Ansari \\ Stanford University, micro1 \\ \texttt{ali@micro1.ai} \and
  Aruj Mahajan \\ micro1 \\ \texttt{aruj@micro1.ai} \and
  Amirhossein Afsharrad \\ Stanford University \\ \texttt{afsharrad@stanford.edu} \and
  Seyed Shahabeddin Mousavi \\ Stanford University \\ \texttt{ssmousav@cs.stanford.edu} 
}
\begin{document}
\date{}
\maketitle

\begin{abstract}
Voice-based conversational AI systems increasingly rely on cascaded architectures that combine speech-to-text (STT), large language models (LLMs), and text-to-speech (TTS) components. We present a large-scale empirical comparison of STT × LLM × TTS stacks using data sampled from over 300,000 AI-conducted job interviews. We used an LLM-as-a-Judge automated evaluation framework to assess conversational quality, technical accuracy, and skill assessment capabilities. Our analysis of five production configurations reveals that a stack combining Google's STT, GPT-4.1, and Cartesia's TTS  outperforms alternatives in both objective quality metrics and user satisfaction scores. Surprisingly, we find that objective quality
metrics correlate weakly with user satisfaction scores, suggesting that user experience in voice-based AI systems depends on factors beyond technical performance. Our findings provide practical guidance for selecting components in multimodal conversations and contribute a validated evaluation methodology for human-AI interactions.
\end{abstract}

\section{Introduction}

Voice-based conversational AI systems are rapidly transforming domains from customer service to healthcare, education, and recruitment. These systems typically employ a cascaded architecture where speech-to-text (STT) engines transcribe user input, large language models (LLMs) generate responses, and text-to-speech (TTS) systems synthesize audio output. While each component has been extensively studied in isolation, understanding how different combinations perform in real-world, high-stakes applications remains an open challenge.

This paper addresses three key research questions:
\begin{enumerate}
    \item \textbf{RQ1:} How do different STT \(\times\) LLM \(\times\) TTS combinations affect conversational quality and technical accuracy in voice-based AI systems?
    \item \textbf{RQ2:} What is the relationship between objective quality metrics and user satisfaction in AI-conducted conversations?
    \item \textbf{RQ3:} Can automated LLM-based evaluation reliably assess the quality of voice-based AI interactions?
\end{enumerate}

To answer these questions, we leverage a unique dataset of over 5,000 AI-conducted job interviews from a production system operating at scale (averaging 1,500 interviews per day). The system conducts 20-30 minute adaptive audio interviews in 15+ languages, able to interview each candidate dynamically, in any niche field, at any level of expertise.

Our main contributions are:
\begin{itemize}
    \item The first large-scale production comparison of five cascaded STT \(\times\) LLM \(\times\) TTS architectures in a real-world application.
    \item A validated evaluation framework using LLM-as-a-Judge for assessing voice-based conversational AI.
    \item Empirical findings revealing the weak correlation between technical quality metrics and user satisfaction.
    \item Analysis of component interdependencies in cascaded voice AI systems.
\end{itemize}

\section{Related Work}

\subsection{Cascaded vs. End-to-End Speech-Language Systems}

Recent research has explored two primary architectures for voice-based conversational AI. The cascaded approach, where STT, LLM, and TTS components operate sequentially, offers modularity and flexibility. Huang et al. demonstrate this in AudioGPT, augmenting ChatGPT with off-the-shelf ASR and TTS modules for multi-round spoken dialogues \cite{huang2023audiogpt}.

In contrast, unified models like SpeechGPT \cite{zhang2023speechgpt} process speech tokens directly, potentially reducing error propagation. These early multimodal LLMs demonstrate end-to-end spoken dialogue generation feasibility, though their half-duplex design and high computational cost mean cascaded setups remain competitive for real-time use.

\subsection{Real-Time Conversational Systems}

To enable more natural conversations, recent work addresses latency and full-duplex interaction. Ma et al. introduce a ``listening while speaking" language model (LSLM) that processes incoming audio while generating speech, achieving full-duplex dialogue via a specialized decoder \cite{ma2024lslm}. Ginart et al. present an asynchronous agent framework with an event-driven architecture running STT and TTS in parallel threads, enabling minimal-delay responses and real-time interruption handling \cite{ginart2024asynchronous}.

\subsection{Evaluation of Conversational AI}

Arora et al. (NAACL 2025) develop ESPnet-SDS, an open-source toolkit standardizing interfaces for different ASR, LLM, and TTS models, enabling controlled benchmarks of pipeline combinations \cite{arora2025espnet}. Their work highlights the need for systematic evaluation, which we address at a large, industrial scale. Our work complements theirs by focusing on in-the-wild production data rather than controlled benchmarks. We also leverage an LLM-as-a-Judge approach, which has been shown to be a scalable and reliable method for evaluating conversational AI systems \cite{zheng2024judging}.

\section{Methodology}

\subsection{System Architecture}
We evaluate a production voice-based interview system developed by micro1 called Zara, with three modular components:
\begin{itemize}
    \item \textbf{Speech-to-Text (STT):} A streaming ASR engine transcribes each utterance in real time as the candidate speaks.
    \item \textbf{Large Language Model (LLM):} The LLM analyzes the latest answer in context of prior turns, required skills from the job description, and any custom questions supplied by the hiring organization. It then composes the next prompt, balancing depth of assessment with conversational flow.
    \item \textbf{Text-to-Speech (TTS):} The composed prompt is sent to a TTS engine, which produces natural-sounding audio so the system can respond immediately and keep the dialogue fluid.
\end{itemize}
Each module can be swapped without altering the others, enabling systematic comparison of different configurations.

\subsection{System Configurations}
We evaluated five distinct STT \(\times\) LLM \(\times\) TTS configurations currently in production, as shown in Table \ref{tab:system_configurations}.

\begin{table}[h!]
\centering
\caption{System Configurations Evaluated in Production}
\label{tab:system_configurations}
\begin{tabular}{llll}
\toprule
\textbf{Config} & \textbf{STT} & \textbf{LLM} & \textbf{TTS} \\
\midrule
C1 & Google STT & GPT-4.1 & OpenAI TTS \\
C2 & Google STT & GPT-4o & OpenAI TTS \\
C3 & Google STT & GPT-4.1 & Cartesia TTS \\
C4 & Whisper STT & GPT-4o & OpenAI TTS \\
C5 & Whisper STT & Groq2 & OpenAI TTS \\
\bottomrule
\end{tabular}
\end{table}

\subsection{Dataset}
Our dataset comprises over 5,000 AI-conducted interviews, sampled and segmented into five unique pairings as described below. Each interview record contains the following:
\begin{itemize}
    \item Complete transcripts with speaker labels
    \item Average duration: 20--30 minutes
    \item Candidate feedback forms and Mean Experience Rating
\end{itemize}
Preprocessing involved grammar normalization using Claude-3.5 Sonnet and removal of filler words to reduce noise.

The dataset also includes a rich set of features such as interview metadata, LLM-as-a-Judge evaluations, and candidate-provided feedback. Key numeric features include the candidate's years of experience, interview duration, and various quality scores from the LLM judge.

\subsection{Evaluation Framework}
We employ a dual evaluation approach, combining direct user feedback with an automated, LLM-based assessment framework validated against human data \cite{aka2025bettertogetherquantifyingbenefits}. Our evaluation is based on four key metrics that are defined in Table \ref{tab:metric_definitions}.

\begin{table}[h!]
\centering
\caption{Definition of Key Performance Metrics}
\label{tab:metric_definitions}
\begin{tabular}{lp{0.65\textwidth}}
\toprule
\textbf{Metric} & \textbf{Definition} \\
\midrule
\textbf{User Rating (1-5)} & Collected from candidates post-interview, this 1-5 star rating (\texttt{rate\_star}) measures their overall satisfaction with the experience. \\
\addlinespace
\textbf{Accuracy} & An LLM-as-a-Judge score (\texttt{llm\_judge\_accuracy\_score}) evaluating if the system's assessment of the candidate's skills aligns with the interview transcript. \\
\addlinespace
\textbf{CQ\textsuperscript{*} Overall} & An LLM-as-a-Judge score for conversational quality (\texttt{llm\_judge\_cq\_overall}), assessing the natural flow, engagement, and etiquette of the dialogue. \\
\addlinespace
\textbf{TQ\textsuperscript{*} Overall} & An LLM-as-a-Judge score for technical quality (\texttt{llm\_judge\_tq\_overall}), assessing the relevance, clarity, and logical progression of interview questions. \\
\bottomrule
\end{tabular}
\vspace{0.5em}
\hspace*{1em}\parbox{0.90\linewidth}{\small\textsuperscript{*} CQ = Conversational Quality, TQ = Technical Quality}
\end{table}

To reduce human grading costs and promote consistency, our automated framework uses Claude 3.5 Sonnet to score each interview transcript. The overall scores for Conversational and Technical Quality are derived from the following sub-metrics:

\begin{itemize}
    \item \textbf{Conversational Quality Sub-metrics:} \emph{Dialogue Flow} (natural progression), \emph{Response Building} (building on candidate answers), and \emph{Acknowledgement} (acknowledging responses).
    \item \textbf{Technical Quality Sub-metrics:} \emph{Skill Alignment} (relevance to the role), \emph{Logical Progression} (how questions build on each other), and \emph{Question Clarity}.
\end{itemize}

The exact prompt for the LLM-based evaluation framework is provided in the Appendix.

\subsection{Statistical Analysis}
We performed a statistical analysis of the collected data.

\begin{enumerate}
    \item \textbf{Descriptive Statistics:} We calculated the mean and standard deviation for each metric across the five configurations.
    \item \textbf{Significance Testing:} We used Levene's test to check for homogeneity of variances. Based on the result, we used either a Welch's ANOVA or a Kruskal-Wallis H test to determine if there were statistically significant differences between the groups.
    \item \textbf{Post-Hoc Analysis:} For metrics with significant differences, we performed a Games-Howell post-hoc test to identify which specific group pairs were different.
    \item \textbf{Correlation Analysis:} We calculated a Pearson correlation matrix to examine the relationships between the various quality metrics and user satisfaction.
\end{enumerate}

\section{Results}

\subsection{Descriptive Statistics}

Table \ref{tab:descriptive_stats} presents the descriptive statistics for key metrics across the five configurations. The Google + GPT-4.1 + Cartesia (C3) configuration consistently achieved the highest mean scores across most LLM-judge metrics, including overall conversational quality (8.78) and technical quality (8.57). This configuration also received the highest average user satisfaction rating (4.53).

\begin{table}[h!]
\centering
\caption{Descriptive Statistics of Key Metrics Across Configurations (Mean \(\pm\) Std. Dev.). CQ is Conversational Quality, TQ is Technical Quality.}
\label{tab:descriptive_stats}
\resizebox{\textwidth}{!}{%
\begin{tabular}{lcccc}
\toprule
\textbf{Configuration} & \textbf{Accuracy} & \textbf{CQ Overall} & \textbf{TQ Overall} & \textbf{User Rating (1-5)} \\
\midrule
Google+GPT-4.1+OpenAI & 7.88 \(\pm\) 1.27 & 8.53 \(\pm\) 0.92 & 8.39 \(\pm\) 1.31 & 4.41 \(\pm\) 0.98 \\
Google+GPT-4o+OpenAI & 8.01 \(\pm\) 0.91 & 8.32 \(\pm\) 0.60 & 8.27 \(\pm\) 0.80 & 4.29 \(\pm\) 0.98 \\
Google+GPT-4.1+Cartesia & 8.12 \(\pm\) 0.55 & \textcolor[rgb]{0,0.5,0}{8.78} \(\pm\) 0.46 & \textcolor[rgb]{0,0.5,0}{8.57} \(\pm\) 0.44 & \textcolor[rgb]{0,0.5,0}{4.53} \(\pm\) 0.84 \\
Whisper+GPT-4o+OpenAI & 8.15 \(\pm\) 0.56 & 8.35 \(\pm\) 0.39 & 8.37 \(\pm\) 0.52 & 4.29 \(\pm\) 1.02 \\
Whisper+Groq2+OpenAI & 8.08 \(\pm\) 0.60 & 8.36 \(\pm\) 0.37 & 8.40 \(\pm\) 0.45 & 4.30 \(\pm\) 0.97 \\
\bottomrule
\end{tabular}%
}
\end{table}

\subsection{Significance Testing}

Our analysis, summarized in Table \ref{tab:significance}, showed statistically significant differences among the configurations for all major metrics, as determined by Welch's ANOVA. This indicates that the choice of components in the STT \(\times\) LLM \(\times\) TTS stack has a material impact on both the perceived quality of the conversation and the technical accuracy of the interview. The only exception was the \texttt{llm\_judge\_soft\_skills\_score} which was not found to be significantly different across groups (p=0.136), though the sample size for this metric was very small (n=10).

Games-Howell post-hoc tests revealed that for \texttt{llm\_judge\_cq\_overall}, C3 was significantly better than all other configurations. For \texttt{llm\_judge\_tq\_overall}, C3 was also significantly better than all other configurations. For \texttt{rate\_star}, C3 was significantly better than C2, C4, and C5.

\begin{table}[h!]
\centering
\caption{Significance Test Results (Welch's ANOVA). All p-values are highly significant, indicating meaningful differences between the system configurations.}
\label{tab:significance}
\begin{tabular}{lccc}
\toprule
Metric & Welch's F-statistic & p-value & $\eta^2$ \\
\midrule
\texttt{llm\_judge\_accuracy\_score} & 12.82 & $<$ .001 & 0.014 \\
\texttt{llm\_judge\_cq\_overall} & 178.23 & $<$ .001 & 0.082 \\
\texttt{llm\_judge\_tq\_overall} & 38.75 & $<$ .001 & 0.015 \\
\texttt{rate\_star} & 10.39 & $<$ .001 & 0.011 \\
\bottomrule
\end{tabular}
\end{table}

\subsection{Analysis}

Table \ref{tab:comprehensive_stats} provides a detailed statistical analysis including Levene's test for homogeneity of variances, Welch's ANOVA results, Kruskal-Wallis test results, and effect sizes for all evaluated metrics. The analysis reveals several important findings:

\begin{itemize}
    \item \textbf{Conversational Quality Metrics:} All CQ metrics show extremely significant differences (p $<$ 0.001) with large effect sizes ($\eta^2$ ranging from 0.041 to 0.086), indicating substantial variation between configurations.
    \item \textbf{Technical Quality Metrics:} TQ metrics also demonstrate highly significant differences, with effect sizes ranging from 0.009 to 0.020, suggesting meaningful but smaller differences compared to conversational quality.
    \item \textbf{Accuracy and User Satisfaction:} Both \texttt{llm\_judge\_accuracy\_score} and \texttt{rate\_star} show significant differences, though with smaller effect sizes ($\eta^2$ = 0.014 and 0.011 respectively).
    \item \textbf{Soft Skills Assessment:} The \texttt{llm\_judge\_soft\_skills\_score} shows no significant differences across groups (p = 0.136), likely due to the small sample size (n = 10).
\end{itemize}

\begin{table}[h!]
\centering
\caption{Analysis Results for All Metrics}
\label{tab:comprehensive_stats}
\begin{tabular}{lcccccc}
\toprule
Metric & Levene's F & Levene's p & Welch's F & Welch's p & $\eta^2$ & Test Used \\
\midrule
\texttt{llm\_judge\_accuracy\_score} & 17.58 & $<$ .001 & 12.82 & $<$ .001 & 0.014 & Welch ANOVA \\
\texttt{llm\_judge\_cq\_overall} & 45.65 & $<$ .001 & 178.23 & $<$ .001 & 0.082 & Welch ANOVA \\
\texttt{llm\_judge\_tq\_overall} & 17.62 & $<$ .001 & 38.75 & $<$ .001 & 0.015 & Welch ANOVA \\
\texttt{llm\_judge\_cq\_dialogue\_flow} & 29.25 & $<$ .001 & 204.58 & $<$ .001 & 0.086 & Welch ANOVA \\
\texttt{llm\_judge\_cq\_response\_building} & 28.86 & $<$ .001 & 189.43 & $<$ .001 & 0.073 & Welch ANOVA \\
\texttt{llm\_judge\_cq\_acknowledgement} & 38.28 & $<$ .001 & 98.63 & $<$ .001 & 0.042 & Welch ANOVA \\
\texttt{llm\_judge\_tq\_skill\_alignment} & 14.08 & $<$ .001 & 19.07 & $<$ .001 & 0.009 & Welch ANOVA \\
\texttt{llm\_judge\_tq\_logical\_progression} & 10.64 & $<$ .001 & 44.66 & $<$ .001 & 0.017 & Welch ANOVA \\
\texttt{llm\_judge\_tq\_question\_clarity} & 28.03 & $<$ .001 & 51.63 & $<$ .001 & 0.020 & Welch ANOVA \\
\texttt{llm\_judge\_soft\_skills\_score} & 0.23 & 0.872 & 0.00 & 1.000 & 0.578 & Kruskal-Wallis \\
\texttt{rate\_star} & 9.22 & $<$ .001 & 10.39 & $<$ .001 & 0.011 & Welch ANOVA \\
\bottomrule
\end{tabular}
\end{table}

\textbf{Note:} $\eta^2$ (eta-squared) represents the effect size, indicating the proportion of variance explained by group differences. Values of 0.01, 0.06, and 0.14 represent small, medium, and large effects respectively.

\subsection{Visualizations of Performance}

\begin{figure}[!ht]
\centering
\includegraphics[width=\textwidth]{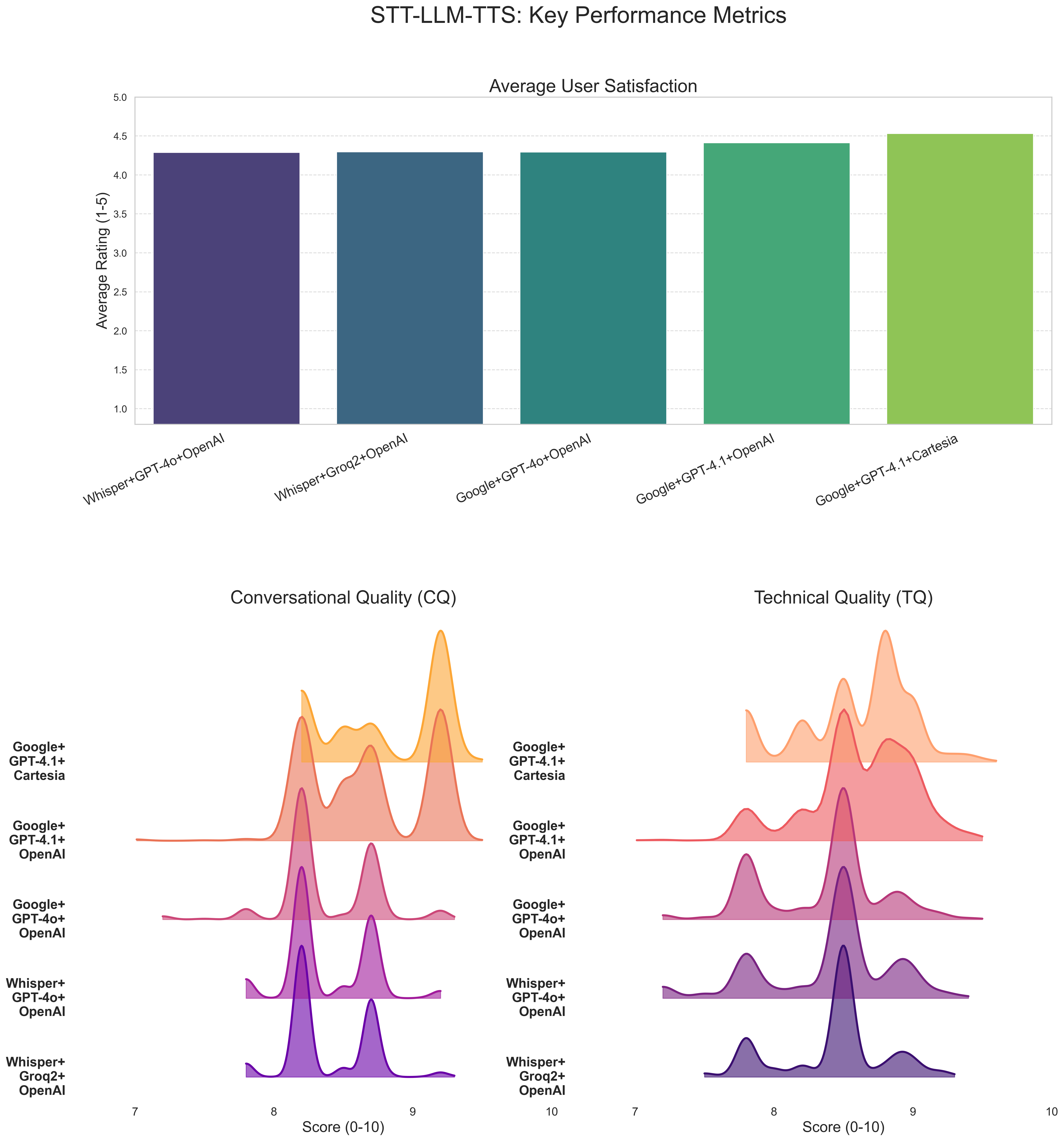}
\caption{Evaluation showing performance of all configurations across key metrics.}
\label{fig:dashboard}
\end{figure}

Figure \ref{fig:dashboard} provides an overview of the performance of each configuration across all metrics.

\begin{figure}[!ht]
\centering
\includegraphics[width=0.8\textwidth]{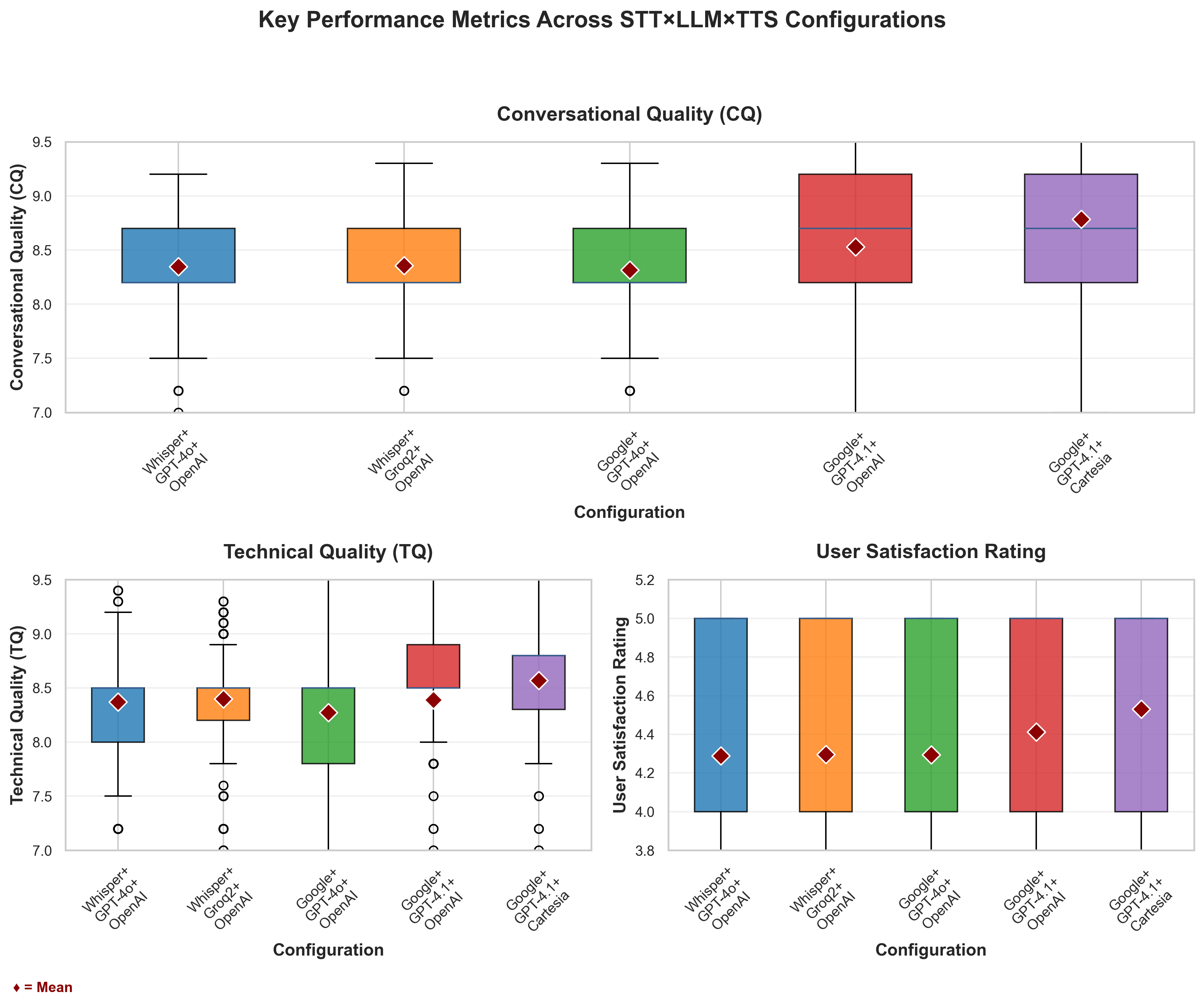}
\caption{Combined boxplot showing Conversational Quality (top), Technical Quality (bottom left), and User Satisfaction Rating (bottom right) across STT × LLM × TTS configurations. Red diamond markers indicate mean values. The Google + GPT-4.1 + Cartesia configuration shows consistently superior performance across all metrics.}
\label{fig:boxplots}
\end{figure}

The combined boxplot visualization in Figure \ref{fig:boxplots} illustrates the distribution of scores for the three key performance metrics across all configurations. The \texttt{Google + GPT-4.1 + Cartesia} configuration (C3) demonstrates superior performance across all metrics, with higher median scores and tighter distributions, indicating more consistent and reliable performance compared to alternative configurations.

\subsection{Component Contribution Analysis}

\begin{figure}[!ht]
\centering
\includegraphics[width=\textwidth]{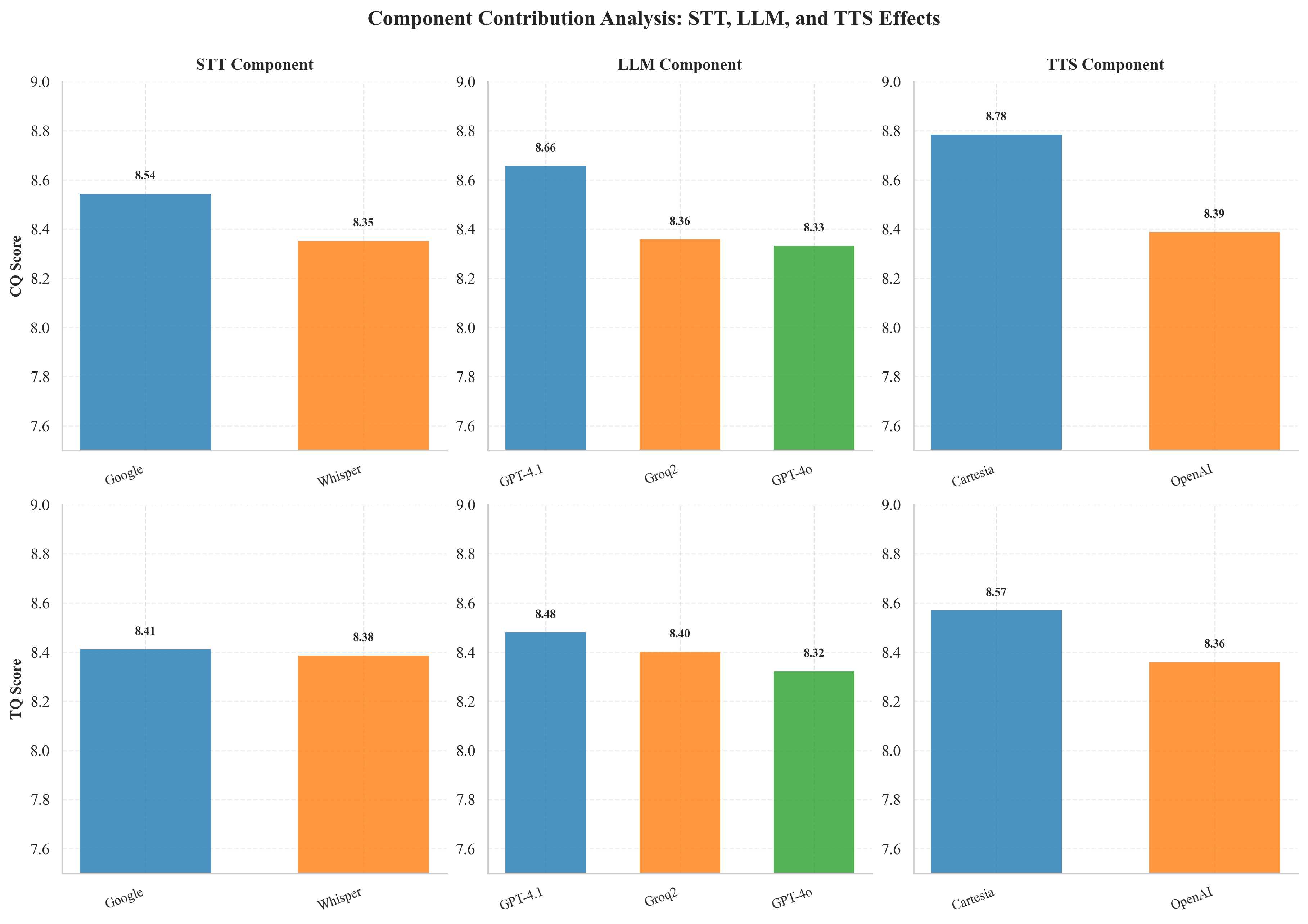}
\caption{Component Contribution Analysis showing individual STT, LLM, and TTS effects across key quality metrics. Each subplot displays performance by component type.}
\label{fig:component_analysis}
\end{figure}

While the previous results demonstrate clear performance differences between configurations, a critical question remains: which individual components (STT, LLM, or TTS) contribute most to these differences? Figure \ref{fig:component_analysis} presents a decomposition of our five configurations to isolate the effects of each component type.

The analysis tells important findings about component contributions:

\textbf{STT has the dominant impact on system performance.} Google STT consistently outperforms Whisper STT across all metrics on the 10-point LLM-as-a-Judge scales. This suggests that STT quality has cascading effects throughout the entire pipeline, as transcription errors propagate to subsequent processing stages. The sample size (n=3,006 for Google) provides confidence in these results.

\textbf{LLM choice shows moderate but consistent effects.} GPT-4.1 outperforms GPT-4o across conversational quality (8.66 vs 8.33), technical quality (8.48 vs 8.32), and user satisfaction (4.53 vs 4.29). Interestingly, GPT-4o appears to represent a middle ground in performance, suggesting that different versions of LLM may optimize for different aspects of conversational quality.

\textbf{TTS contributes smaller but meaningful improvements.} Cartesia TTS consistently outperforms OpenAI TTS with improvements of 0.1-0.2 points across all metrics. While these effects are smaller than STT or LLM contributions, they are systematic and may be particularly important for user experience in voice-based interactions.

These findings have important implications for system design and resource allocation. The dominance of STT effects suggests that organizations should prioritize STT quality when optimizing cascaded voice AI systems, as improvements here yield the largest performance gains. The moderate LLM effects indicate that model selection remains important but secondary to transcription quality. Finally, the consistent TTS effects, while smaller, may be important for user satisfaction in production deployments where voice naturalness directly impacts user experience.

\subsection{Correlation Analysis}

\begin{figure}[!ht]
\centering
\includegraphics[width=.6\textwidth]{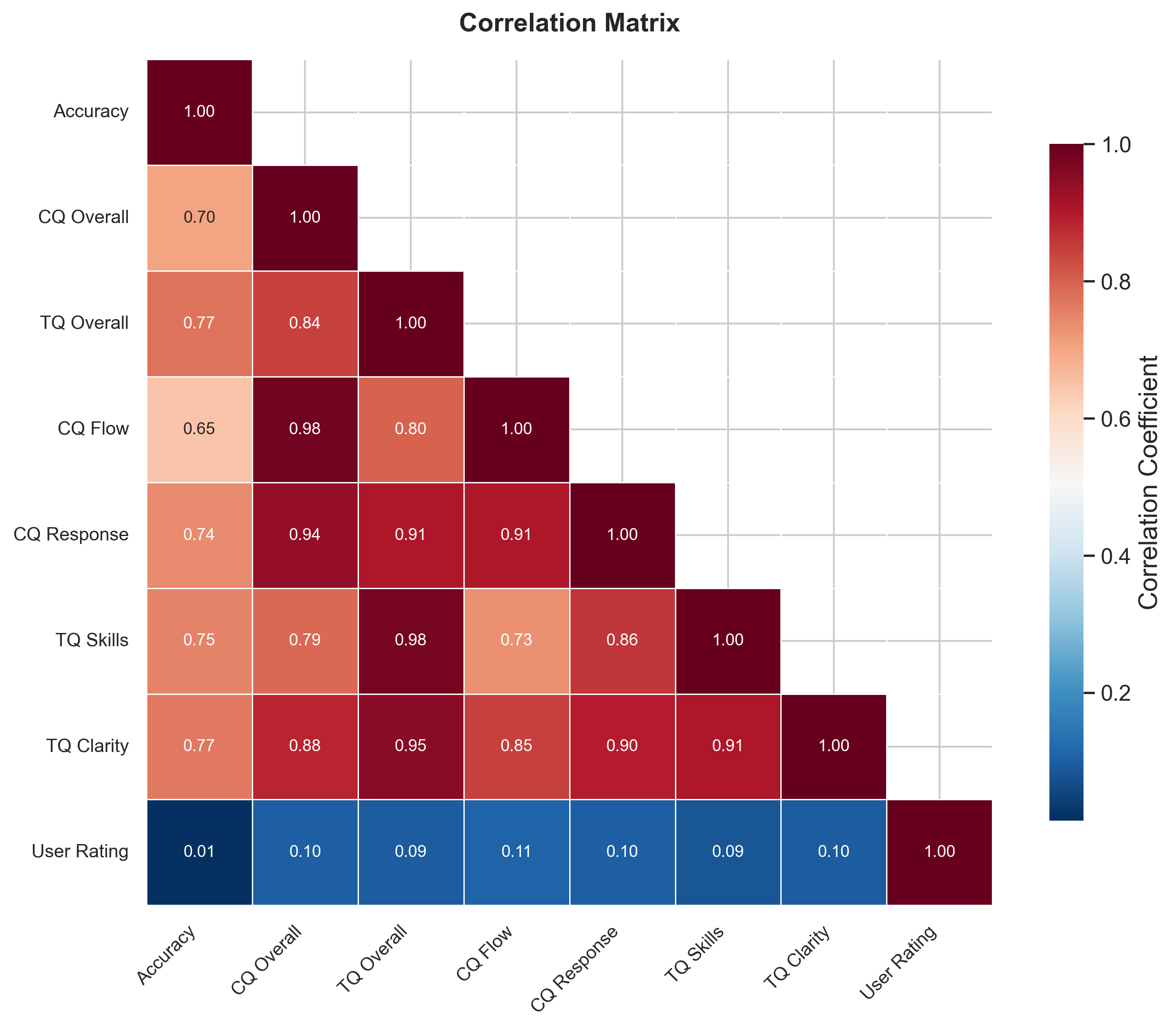}
\caption{Correlation matrix showing relationships between LLM-as-a-Judge metrics and user satisfaction. The upper triangle is masked for clarity. Note the consistently weak correlations between LLM-judge metrics and user experience ratings, signaling a possible disconnect between automated quality assessments and actual user experience.}
\label{fig:correlation}
\end{figure}

We found a weak correlation between the objective LLM-as-a-Judge metrics and the subjective user ratings. Figure \ref{fig:correlation} shows the key correlations in our evaluation. Although the different LLM-as-a-Judge metrics are highly correlated with each other (e.g., CQ Overall and TQ Overall show strong positive correlations), their correlation with user ratings is consistently low. Most LLM-as-a-Judge metrics show correlations with user satisfaction below 0.11, pointing to a possible disconnect between automated quality assessments and actual user experience.

\section{Discussion}

Our results provide several key insights. First, the choice of components in a cascaded voice AI system has a significant impact on performance. The \texttt{Google + GPT-4.1 + Cartesia} configuration emerged as the clear winner in our tests, excelling in both conversational and technical aspects. Importantly, our component contribution analysis reveals that this superior performance is driven primarily by the STT component, with moderate contributions from the LLM and smaller but consistent improvements from the TTS component.

Second, the weak correlation between our automated LLM-as-a-Judge metrics and human user satisfaction ratings is a critical finding that validates the importance of our dual evaluation approach. While our automated LLM-as-a-Judge framework provides consistent, cost-effective assessment of Question Quality, Conversational Flow, and Answer Assessment, it appears to miss key aspects of user experience that drive satisfaction. This suggests that users may value factors beyond conversational performance, such as perceived empathy, voice naturalness, or overall interaction design. The moderate correlation with soft skills assessment (0.53) indicates that users' perception of the AI's interpersonal qualities may be more important than conversational accuracy. This finding underscores the value of collecting direct human feedback alongside automated metrics, as each captures different dimensions of system performance.

Third, the LLM-as-a-Judge framework proved to be a scalable and effective way to evaluate our system. The high correlation between the sub-metrics suggests that the judge model is consistent in its evaluations.

\subsection{Limitations}
Several limitations should be considered when interpreting our results:

\textbf{Temporal Confounds:} The non-overlapping deployment periods of different configurations introduce potential confounds from seasonal variations, candidate pool changes, or system improvements over time. Unobserved temporal effects may influence our performance comparisons.

\textbf{Limited TTS Variety:} Our study design prevented comprehensive assessment of TTS impact on overall performance, as most configurations used the same TTS component. Although our component contribution analysis shows TTS effects through the Cartesia vs. OpenAI comparison, a full factorial design with multiple TTS options across all configurations would provide more robust insights into TTS contributions.

\textbf{Domain Specificity:} Our study focused specifically on job interview scenarios, which may limit generalizability to other conversational AI applications with different interaction patterns, success criteria, or user expectations. The evaluation metrics and user satisfaction patterns observed may not transfer directly to domains such as customer service, education, or healthcare applications.

\textbf{Self-Selection Bias:} User satisfaction ratings suffer from potential self-selection bias, as not all candidates provided feedback. Candidates who had particularly positive or negative experiences may be more likely to respond, potentially skewing satisfaction scores and affecting the generalizability of user experience findings.

\subsection{Future Work}
This study provides a foundation for the evaluation of voice-based AI interview systems using combined STT, LLM, and TTS stacks. Future work will expand the empirical comparison as new models and synthesis engines are introduced, enabling an ongoing assessment of performance across conversational and technical quality along with user satisfaction.

A promising direction to improve the reliability of the evaluation involves using multiple LLM judges and aggregating their scores. Although our current approach uses a single Claude 3.5 Sonnet judge, ensemble methods combining multiple LLM judges (e.g., GPT-4, Claude, and other advanced models) could provide more robust and consistent evaluations. Score aggregation techniques, such as weighted averaging or consensus-based scoring, could reduce individual judge biases and improve the overall reliability of automated quality assessments. This approach would be particularly valuable given the weak correlation we observed between automated metrics and user satisfaction, as a more diverse set of judges might better capture the nuanced aspects of conversational quality that drive user experience.

\section{Conclusion}

We conducted a large-scale comparison of five different STT \(\times\) LLM \(\times\) TTS stacks in a production AI interview system. Our findings show that the \texttt{Google STT + GPT-4.1 + Cartesia TTS} configuration significantly outperforms the others on both objective metrics and user satisfaction. More importantly, we confirmed that objective metrics of conversational and technical quality have a surprisingly weak correlation with user satisfaction. This highlights the need for a more holistic approach to evaluating voice-based AI systems, one that goes beyond conversational performance to include aspects of user experience and perceived interaction quality. Our work provides practical guidance for building and evaluating the next generation of conversational AI.

\bibliographystyle{plain}
\bibliography{references}

\newpage
\appendix
\section{LLM-as-a-Judge System Prompt}
The following prompt was used with the `claude-3-5-sonnet-20241022` model to evaluate each interview transcript.

\begin{verbatim}
You are tasked with **scoring the evaluations** provided for an interview based on the 
**interview transcript**.

### **Input Data:**
- index: "{index}"
- LLM model: {model}
- interview_transcript: "{interview_transcript}"
- evaluation_result: "{evaluation_result}" **(Ensure "Mid-level" is replaced with "Experienced")**
- soft_skills_evaluation_result: "{soft_skills_evaluation_result}"

---
### **Step 1: Analyze the results coming from **evaluation_result** and give a score out of 10
based on the interview transcript and the LLM model.

### **Step 2: Evaluate the softskills evaluation coming from **soft_skills_evaluation_result**
and give a score out of 10 based on the interview transcript.
Soft skills were evaluated based on **Common European Framework of Reference (CEFR) 
language assessment**, focusing on:
- **Language Fluency & Coherence**
- **Vocabulary Range & Accuracy**
- **Grammatical Competence**
- **Critical Thinking & Logical Argumentation**
- **Topic Engagement & Responsiveness**

#### **CEFR Level Mapping for Soft Skills:**
- **A1**: Beginner
- **A2**: Elementary
- **B1**: Intermediate
- **B2**: Upper-Intermediate
- **C1**: Advanced
- **C2**: Near-Native

### **Step 3: Evaluate the interview quality based on the transcript**
Evaluate the interview across multiple dimensions, giving scores only based on 
interviewer's responses not candidates. For each criterion, provide a score from 
1-10 with one decimal place precision.

Scoring criteria for interview quality:

Conversational Quality:
- dialogue_flow: How natural and engaging is the dialogue flow by the interviewer
- response_building: How well interviewer builds upon candidate responses and 
                     his professional etiquettes
- acknowledgement: How well interviewer acknowledges candidate responses

Technical Quality:
- skill_alignment: Alignment of questions with stated skill requirements
- logical_progression: How well questions build to assess breadth and depth
- question_clarity: Clarity and unambiguity of questions

---
### **Your Task**
1. **Score the Evaluations**:
   - **Accuracy**: Does the evaluation align with the transcript and correctly reflect 
                  the candidate's performance?
2. **Evaluate Interview Quality**:
   - Score both conversational and technical aspects of the interview

---
### Ensure Ratings Are Standardized**
- **Replace all occurrences of "Mid-level" (case-insensitive) with "Experienced"** 
  in `evaluation_result`.
- Ensure that all **evaluations follow the correct rating format** 
  (`Not Experienced, Junior, Experienced, Senior`).

---
### **Output:**
Return a **dictionary** containing the following keys. 
**Do not include any other keys or any explanation like I'll analyze the interview 
and provide the requested scoring.**:

```json
{{
    "index_from_chain": "{index}",
    "interview_type_from_parallel_chain": "<Technical or Non-Technical>",
    "accuracy_score_from_baseline": "<0-10>",
    "accuracy_score_from_baseline_explanation": "<Brief explanation for assigned scores>",
    "conversational_quality_overall_score": <score>,
    "technical_quality_overall_score": <score>,
    "conversational_quality_dialogue_flow": <score>,
    "conversational_quality_response_building": <score>,
    "conversational_quality_acknowledgement": <score>,
    "technical_quality_skill_alignment": <score>,
    "technical_quality_logical_progression": <score>,
    "technical_quality_question_clarity": <score>
}}
```
\end{verbatim}

\end{document}